\documentclass[journal=jpccck,layout=twocolumn]{achemso}
\usepackage{graphicx}  % include pdf figure files
\usepackage{todonotes}
\usepackage{booktabs}  % pretty tables
\newcommand{\ra}[1]{\renewcommand{\arraystretch}{#1}}  % pretty table spacing
\usepackage{microtype}  % prettier typesetting
\usepackage{nameref}

\title{Development of a coarse-grained water forcefield via
multistate iterative Boltzmann inversion}
\author{Timothy C. Moore}
\affiliation{Department of Chemical and Biomolecular Engineering, Vanderbilt University, Nashville, TN 37235}
\alsoaffiliation{Vanderbilt University Center for Multiscale Modeling and Simulation (MuMS), Nashville, TN 37235}
\author{Christopher R. Iacovella}%
\affiliation{Department of Chemical and Biomolecular Engineering, Vanderbilt University, Nashville, TN 37235}
\alsoaffiliation{Vanderbilt University Center for Multiscale Modeling and Simulation (MuMS), Nashville, TN 37235}
\author{Clare McCabe}%
\affiliation{Department of Chemical and Biomolecular Engineering, Vanderbilt University, Nashville, TN 37235}
\alsoaffiliation{Vanderbilt University Center for Multiscale Modeling and Simulation (MuMS), Nashville, TN 37235}
\alsoaffiliation{Department of Chemistry, Vanderbilt University, Nashville, TN 37235}
\email{c.mccabe@vanderbilt.edu}
\begin{document}

\maketitle
\section{Introduction}
Coarse-grained (CG) models have proven to be useful in many fields of chemical
research
\cite{wang2005unique,bhargava2007nanoscale,karimi2010studying,padding2002time,
harmandaris2010temperature,sun2006crossover,milano2005mapping,
shinoda2008coarse,lee2011coarse,srinivas2004self},
allowing molecular simulation to be performed on larger system sizes and access
longer timescales than is possible with atomistic-level models, enabling complex
phenomena such as hierarchical self-assembly to be described.
\cite{Nguyen2004,IacovellaPNAS2011}
In CG simulations of aqueous systems, especially ones with significant amounts
of hydrophobic and/or hydrophilic interactions, the water model is important and
can have a major impact on the resulting properties of the
system.\cite{Hadley2012}

While the assignment of atoms to CG beads (i.e., defining the CG mapping) is
relatively straightforward for most chemical systems (e.g., aggregating four
methyl groups bonded in sequence into a single CG bead), mapping an atomistic
water trajectory to the CG level (i.e., grouping several water molecules into a
single CG bead) is not as well-defined given the lack of permanent bonds between
water molecules.
This ambiguity presents a problem for structure-based methods that require an
atomistic configuration to be mapped to the corresponding CG configuration,
e.g., to generate target radial distribution functions (RDFs) against which the
forcefield is optimized.
As such, the majority of many-to-one CG models of water (i.e., where one CG bead
represents multiple water molecules) have instead been derived by assuming a
functional form of the forcefield and optimizing the associated parameters to
match selected physical properties of water, such as density, vaporization
enthalpy, surface tension, etc.\cite{Hadley2012,Basdevant2004,Basdevant2006,
Masella2011,marrink2007martini,Chiu2010,Shinoda2007}
For example, Chiu \textit{et al.} developed a 4:1 CG water forcefield by
optimizing the parameters of a Morse potential to accurately reproduce the
surface tension and density of liquid water.\cite{Chiu2010}
Despite capturing the interfacial properties and density, this potential
overestimates structural correlations, as one might expect given that structural
data was not used in the optimization.

Recently, Hadley and McCabe\cite{hadley2010on} proposed a method for mapping
configurations of atomistic water to their CG representations using the
$k$-means clustering algorithm. 
Subsequently in related work, van Hoof \textit{et al.}\cite{VanHoof2011a}
developed the CUMULUS method for mapping atoms to CG beads.
Both methods enable dynamic mapping of multiple water molecules to a single CG
bead, allowing structure-based schemes to be used.
Here, dynamic refers to a CG mapping that changes over the course of the
atomistic trajectory, i.e., different water molecules are assigned to different
CG beads in each frame of the atomistic trajectory.
Both works employed the iterative Boltzmann inversion
(IBI)\cite{reith2003deriving} method to derive the intermolecular interactions
by optimizing a numerical, rather than analytical, potential to reproduce RDFs
calculated from the atomistic-to-CG mapped
configurations.\cite{VanHoof2011a,hadley2010on}
The forcefields derived are similar and show good agreement with the structural
properties and density of the atomistic water models studied.
However, neither model is able to accurately reproduce interfacial properties,
since they were derived solely from bulk fluid data.
This failure to capture the interfacial properties is a consequence of the
single-state nature of the IBI approach, and may alter the balance of
hydrophobic and hydrophilic interactions when using these water models in
multicomponent systems.

Recently, the multistate IBI (MS IBI) method was developed as an extension of
the original IBI approach,\cite{moore2014derivation} with the goal of reducing
state dependence and structural artifacts often found in IBI-based potentials.
\cite{hadley2010coarse,bayramoglu2012coarse,qian2008temperature}
A significant issue related to the IBI method is that a multitude of potentials
may give rise to similar RDFs, and the method cannot necessarily differentiate
which of the many potentials is most accurate, as only RDF matching is
considered.
MS IBI operates based on the idea that different thermodynamic states will
occupy different regions of potential ``phase space'' (i.e., regions where
potentials give rise to similar RDFs), and that the most transferable, and thus
most accurate, potential lies in the overlap of phase space for the different
states.
That is, by optimizing a potential simultaneously against multiple thermodynamic
states, MS IBI provides constraints to the optimization, forcing the method to
derive potentials that exist in this overlap region, and thus are transferable
among the states considered.
The MS IBI approach has been shown to reduce state dependence and improve the
quality of the derived potentials, as compared to the original IBI
method.\cite{moore2014derivation}

In this work, multistate iterative Boltzmann inversion (MS IBI) is used to
derive an intermolecular potential that captures both bulk and interfacial
properties of water, improving upon the CG water model of Hadley and
McCabe.\cite{hadley2010on}
Again, optimizations are carried out using the MS IBI method, where both bulk
and interfacial systems are used simultaneously as target conditions for the
optimization.
MS IBI is also used, for the first time, in a multi-ensemble context, enabling
optimizations in both the canonical (NVT) and isothermal-isobaric (NPT)
ensembles to be performed simultaneously to derive the density-pressure
relationship of the system. 
To further constrain the optimization, a slightly modified version of the Chiu
\textit{et al.} CG water forcefield, optimized for surface tension, is used as a
starting condition, allowing the MS IBI method to make specific modifications to
the potential to improve structural properties.
The remainder of the paper is organized as follows: In Methods, a brief overview
of the $k$-means clustering and MS IBI algorithms is given and the models used
are described.
The potential derivation is then presented, validated, and compared to existing
CG water models in the Results section and finally, conclusions are drawn about
the applicability of the derived CG model and the broader applicability of the
MS IBI method discussed.

\section{Methods \label{sec:methods}}
\subsection{$k$-means Clustering Algorithm}
Mapping a water trajectory to the CG level is inherently different than mapping
a larger molecule's trajectory, since for water, atoms mapped to a single CG bead
exist on different molecules.
Furthermore, the water molecules mapped to a common bead are not likely to
remain associated throughout the full simulation because of thermal diffusion.
A dynamic mapping scheme is therefore required for water.
Following the work of Hadley and McCabe,\cite{hadley2010on} the $k$-means
algorithm is used to map the atomistic water trajectory to the CG level. 
In short, $k$-means is a clustering algorithm that is used to find clusters of
data points in a large data set.
The positions of the water molecules are here analogous to the points in the
data set and waters mapped to a single bead are analogous to the clusters.
More details of the algorithm can be found
elsewhere.\cite{hadley2010on,hartigan1979algorithm}
While the $k$-means algorithm can be used to group together any number of 
water molecules, a 4:1 mapping is chosen, as this was found in prior to provide the best
balance between accuracy and computational efficiency,\cite{hadley2010on} and
4:1 models are common in the
literature.\cite{marrink2007martini,Chiu2010,hadley2010on}

\subsection{Multistate Iterative Boltzmann Inversion Method
  \label{subsec:msibi}}
MS IBI was used to derive the intermolecular potential between water beads.  
The goal of MS IBI is to derive a single potential that can be used over a range
of thermodynamic states.
As an extension of the original IBI method,\cite{reith2003deriving} the potential
is updated based on the average differences in CG and target RDFs at multiple states
(i.e., a single potential for each pair is updated based on RDFs from multiple
states).
The potential is adjusted according to
\begin{equation}
  V_{i+1}(r) = 
  V_{i}(r) - \frac{1}{N}\sum_s \alpha_s(r) k_B T_s
  \ln \left[ \frac{g^*_s(r)}{g^i_s(r)} \right],
  \label{eq:msibi}
\end{equation}
where $V_{i}(r)$ is the pair potential as a function of
separation $r$ at the $i^{th}$ iteration;
$N$ the number of target states;
$\alpha(r)$ an effective weighting factor for each state, allowing more or less
emphasis to be put on a particular target state;
$k_B$ the Boltzmann constant and $T$ the absolute temperature; 
$g^i_s(r)$ the RDF from the CG simulation at iteration $i$ and state $s$;
$g^*_s(r)$ the target RDF at state $s$.
$\alpha_s(r)$ was chosen to be a linear function of the form 
\begin{equation}
\alpha_s(r) = \alpha_{0,s}(1 - r/r_{cut})
  \label{eq:alpha}
\end{equation}
such that $\alpha(r_{cut})=0$, and the potential remains 0 for $r \ge r_{cut}$.
This form of $\alpha(r)$ also places more emphasis on the short-ranged part of
the potential to suppress long-range structural artifacts.

An initial CG potential is assumed for each pair interaction.
In theory, there are no restrictions on the initial potential, so it may take
any form; however, in practice, the initial potential is often the potential of
mean force (PMF) calculated from the Boltzmann inverted RDF.
In this work, rather than taking an average of the PMFs over the states used, the initial
potential used was slightly modified from the water model of Chiu \textit{et
al.}, as discussed below.

A CG simulation is then run with the initial potential.
Based on the RDFs from the CG simulation, the potential is updated according to
Equation~\ref{eq:msibi}.
The updated potential is used as input to the next cycle, and the process is
repeated until some stopping criterion is met. 
Here, the stopping criterion is determined using the following fitness function
\begin{equation}
  f_{fit} = 1 - 
  \frac{\int_0^{r_{cut}} \mathrm{d}r \left| g^i(r) - g^*(r)\right|}
  {\int_0^{r_{cut}} \mathrm{d}r \left| g^i(r)\right| + \left| g^*(r)\right|}
  \label{eq:f_fit}
\end{equation}
where the optimization is stopped when the value of $f_{fit}$ exceeds a
specified value (i.e., meets some tolerance), given below.

\subsection{Models}
Atomistic simulations of pure water were performed with the TIP3P
model.\cite{jorgensen1983comparison}
All atomistic systems contained 5,832 water molecules and were simulated in
LAMMPS\cite{plimpton1995fast,lammpsweb} using a 1~fs timestep.
A cutoff distance of 12 \AA{} was used for the van der Waals interactions; long-range
electrostatics were handled with the PPPM method with a 12~\AA{} real space
cutoff. 
Three distinct states were simulated: bulk, NVT at 1.0~g/mL and 305~K; bulk, NPT
at 305~K and 1.0~atm; and an NVT droplet state at 305~K, where the box from the
bulk NVT state was expanded by a factor of 3 in one direction.
Each atomistic simulation was run for 7~ns.
The atomistic trajectories were mapped to the CG level using the $k$-means
algorithm.
Target RDFs were calculated from the final 5~ns of the mapped trajectory from
each state (bulk NVT, bulk NPT, and droplet NVT).
MS IBI was performed using the target data from each of the three states. 
The initial guess of the potential is given as a Morse potential of the form
\begin{equation}
  V(r) = D_e\left(e^{-2\beta(r-r_{eq})} - 2e^{-\beta(r-r_{eq})} \right)
  \label{eq:morse}
\end{equation}
where $r_{eq}$ is the location of the potential minimum, $-D_e$ is the value of
the potential minimum, and $\beta$ is related to the width of the potential
well.
Parameters are taken to be those from Chiu \textit{et al.}:
$D_e=0.813$~kcal/mol, $\beta=0.556$~\AA{}$^-1$, and $r_{eq}=6.29$~\AA{},
however, we note that the potential was adjusted so that
$\beta=0.5$~\AA{}$^{-1}$ for $r < r_{eq}$. 
This change was made to increase sampling as small separations, because
numerical issues arise in the potential update when the CG RDF is zero but the
target RDF is nonzero.
This modification of the potential will slightly alter the properties as
compared to the original model, as discussed below.
The potential update scaling factor $\alpha_{0,s}$ (see Equations~\ref{eq:msibi}
and \ref{eq:alpha}) was set to 0.7 to avoid large updates to the potential.
The optimizations were stopped when $f_{fit} \ge 0.98$ for each state and
$f_{fit}(i) - f_{fit}(i-1) < 0.001$.

All optimizations were performed with the open-source MSIBI Python
package we developed,\cite{msibiweb} which calls
HOOMD-Blue\cite{anderson2008general,glaser2015strong,hoomdweb} to run the CG
simulations and uses MDTraj\cite{mcgibbon2014mdtraj,mdtrajweb} for RDF
calculations and file-handling.
CG simulations were run at the same states as the atomistic systems.
Initial CG configurations were generated from the CG-mapped atomistic
trajectories at each state.
As a result of the 4:1 mapping, CG water simulations contained 1,458 water
beads.
All CG simulations were run with a 10~fs timestep.
The derived CG potential was set to 0 beyond the cutoff of 12~\AA{}. 

The surface tension $\gamma$ of the droplet state was calculated as 
\begin{equation}
  \gamma = \frac{1}{2}L_z\left< P_{zz} - \frac{P_{xx} + P_{yy}}{2} \right>,
  \label{eq:surface-tension}
\end{equation}
where $L_z$ is the length of the box in the expanded direction, $P_{zz}$ is the
pressure component in the direction normal to the liquid-vapor interfaces,
$P_{xx}$ and $P_{yy}$ are the pressure components in the directions lateral to
the interfaces, and the angle brackets denote a time average.
The factor of $1/2$ is included to account for the two interfaces that are
present in the droplet simulation setup.

\section{Results and Discussion} \label{sec:results}
\subsection{Modified Chiu Potential}
\begin{figure}[h!]
  \centering
  \includegraphics{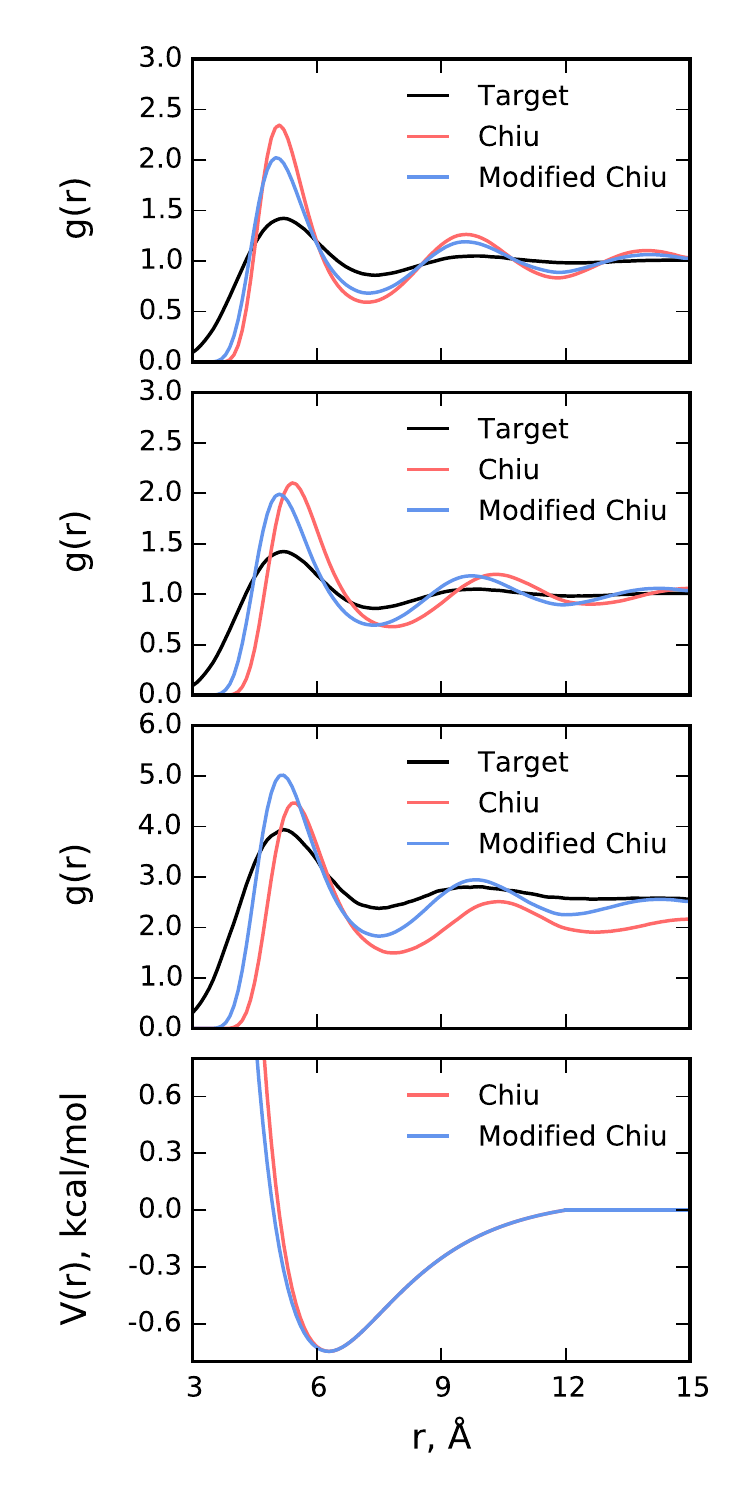}
  \caption{RDFs from simulations using the original and modified Chiu
  potentials. Top: NVT; middle: NPT; bottom-middle: interface; bottom:
comparison of the two potentials.}
  \label{fig:chiu-rdfs}
\end{figure}

We first consider the impact of modifying the Chiu \textit{et al.} potential to
create a softer repulsion.
Figure~\ref{fig:chiu-rdfs} plots the RDFs of the three target states for the
original and modified potentials and the RDF of the 4:1 mapped state (i.e, the
target data used later for the MS IBI optimization). 
The peak location of the NVT state is relatively unchanged; however, upon
modification, there is a slight shift in the first peak for the NPT and
interfacial states, allowing the model to access smaller separations, as was
intended.
The softer potential allows closer contact and thus allows the MS IBI algorithm
to modify this region of the potential where the 4:1 mapped atomistic water has
non-zero values of the RDF. 
The density predicted with both potentials is the same (0.991~$\pm$~0.003~g/mL),
however, due to the softening of the potential, the calculated surface tension
of the droplet changes from 70.3~mN/m to 45~mN/m after the modification,
although we note this value is still sufficient for the droplet to maintain a
stable interface.

\subsection{Potential Derivation and Validation of Bulk Water}
\begin{figure}[h!]
  \centering
  \includegraphics{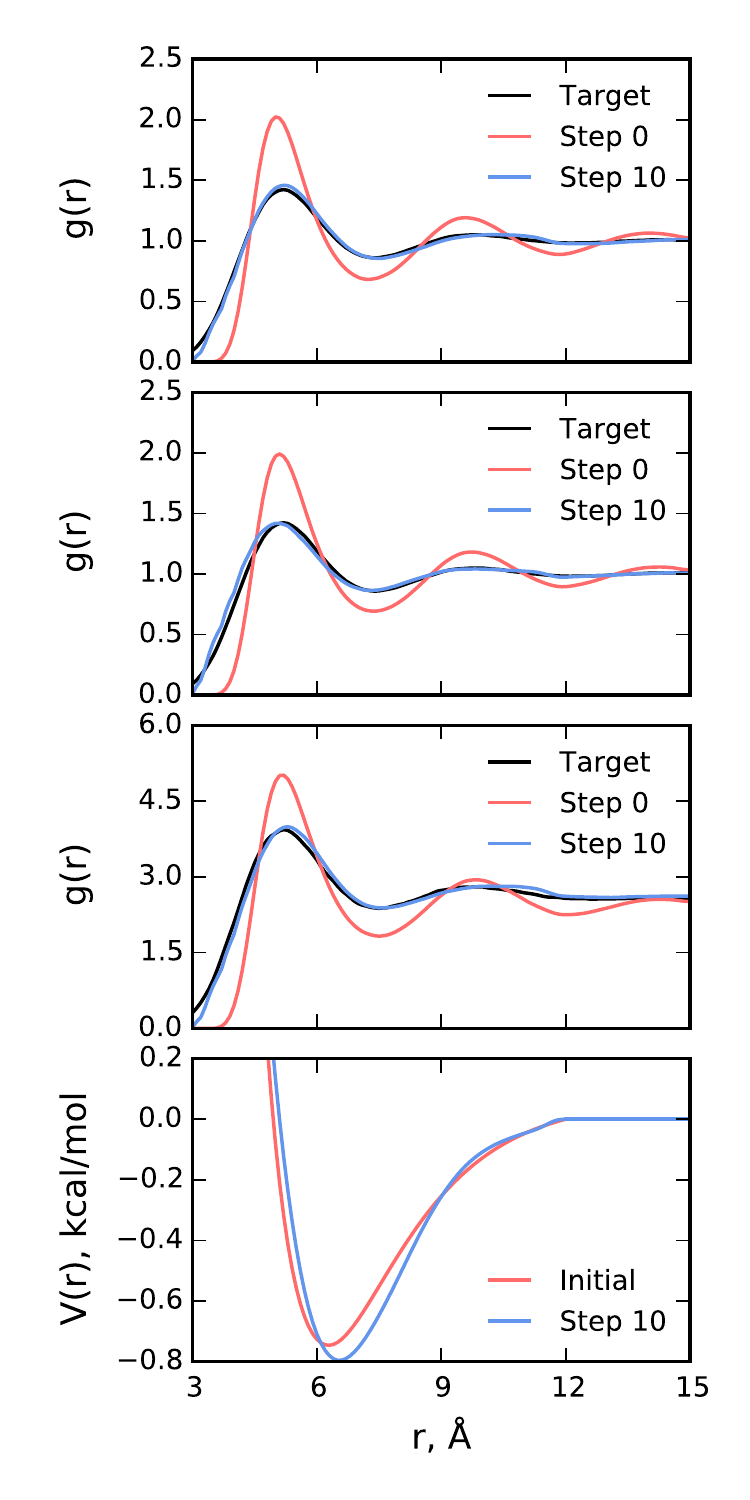}
  \caption{RDFs and potentials from the MS IBI potential derivation. Top: NVT;
  middle-top: NPT; middle-bottom: interface; bottom: potentials. The initial
potential shows significant structural correlations missing from the target
data. The derived potential at ten iterations shows excellent structural
agreement with the target.}
  \label{fig:msibi-results-from-chiu}
\end{figure}

\begin{figure}[ht!]
  \centering
  \includegraphics{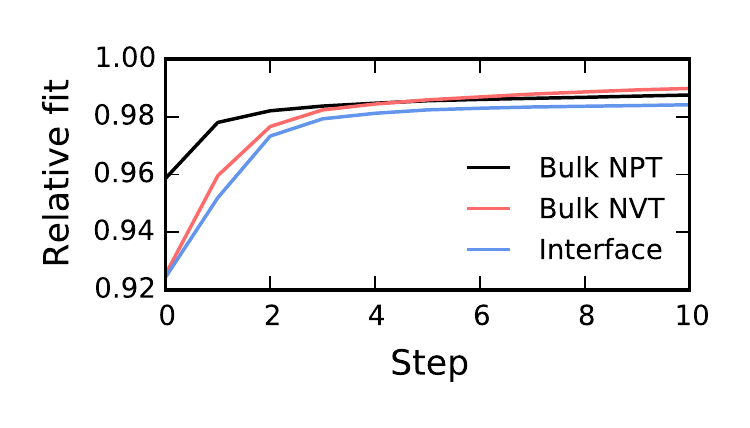}
  \caption{$f_{fit}$ from Equation~\ref{eq:f_fit} as a function of iteration in
the potential derivation. Convergence with the criterion is found after 10
iterations.}
  \label{fig:f-fits}
\end{figure}

Starting from the modified Morse potential of Chiu \textit{et al.}, the new
water forcefield is optimized using the bulk NVT and NPT states and the
interfacial state.
The results of the potential derivation are summarized in
Figure~\ref{fig:msibi-results-from-chiu}, where it is clear that the modified
Chiu \textit{et al.} potentials (i.e., step 0) overestimates the structural
correlations, as was also seen in Figure~\ref{fig:chiu-rdfs} for both the
modified and original potentials.
After only a few iterations, the RDFs match the targets with a high degree of
accuracy. 
This trend is shown in Figure~\ref{fig:f-fits}, which plots the fitness value
from Equation~\ref{eq:f_fit} as a function of iteration.
The value of $f_{fit}$ changes most rapidly in the first 3 steps of the
optimization. 
After 10 iterations, the stopping criteria are met and the optimization stopped.
While the changes to the potential are small, there is a noticeable shift in the
location of the minimum to a slightly larger $r$ value and the potential becomes
slightly more attractive.
Although the shape of the attractive well is mostly unchanged, the potential
more rapidly decays to 0 than the original Morse potential at larger $r$ values,
while the shape of the repulsive regime at small $r$ is changed slightly.
These subtle changes to the potential are sufficient to create significant
changes in the RDF and provide excellent convergence of the structural
correlations.
Note that in Figure~\ref{fig:chiu-rdfs} and \ref{fig:msibi-results-from-chiu} the
RDFs from the interfacial state do not decay to 1 at large $r$.
This is due to the fact that 2/3 of the box is essentially devoid of particles,
but the RDF is normalized based on the volume of the whole simulation box.
This has no effect on the potential update scheme, as both the target and CG
RDFs are normalized by the same factor, which cancels out in
Equation~\ref{eq:msibi}.

In addition to accurately capturing the RDFs, the multi-ensemble approach
provides an accurate estimate of the density at 305~K and 1~atm. NPT simulations
performed using the optimized CG forcefield find a density of
1.027~$\pm$~0.006~g/mL, compared to 1.037~$\pm$0.004~g/mL for TIP3P water.
This approach is successful because the RDFs will not match if the
pressure-density relationship is not satisfied, as the density is implicitly
represented in Equation~\ref{eq:msibi} through the RDF terms (i.e., the RDFs at
the NPT state will not match the target RDFs if the density is significantly
different than the density of the target state).
In contrast, the original IBI method proposed the use of a pressure correction
term of the form $\Delta V(r) =A(1-r/r_{cut})$ to account for the
pressure.\cite{reith2003deriving}
This approach has been successful, but requires a somewhat arbitrary estimate of
the parameter $A$. 
While a method exists for estimating $A$ based on the virial
expression,\cite{wang2009comparative} some degree of trial-and-error is still
necessary.
Furthermore, the multi-ensemble approach within MS IBI does not require direct
calculation of the pressure, which often demonstrates considerable fluctuations,
providing a simpler route to account for the pressure in the CG model.

Calculation of the surface tension of the derived MS IBI potential yields a
value of 42~mN/m, lower than the original Chiu \textit{et al.} potential which
was optimized to match experiment, but only slight perturbed from the modified
potential (45~mN/m).
This reduction is surface tension appears directly related to the softening of
the potential, although, we note that this softening is required to provide an
accurate match of the structure.

\subsection{Validation and Comparison to Other Models}
\begin{figure}[h!]
  \centering
  \includegraphics{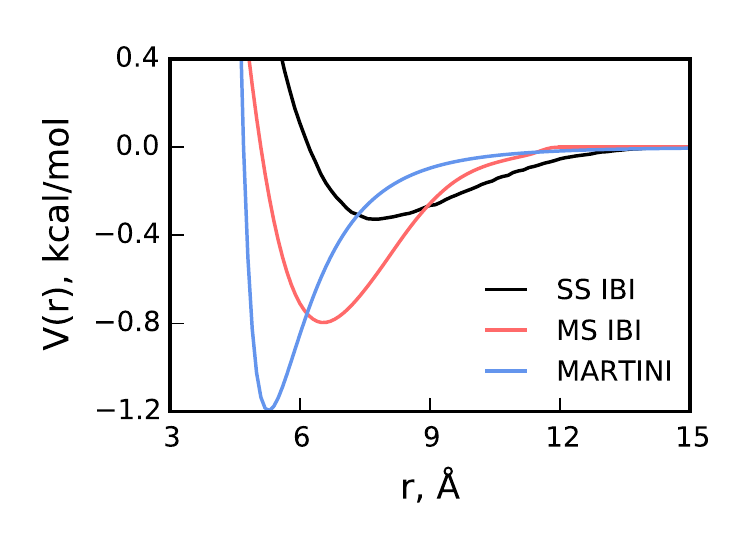}
  \caption{Interaction potentials from the CG water models compared in this
  work. The MS IBI and SS IBI potentials are numerical, derived with
structure-based methods. MARTINI is a Lennard-Jones 12-6 potential.}
  \label{fig:compare-potentials}
\end{figure}

To further explore the efficacy of the MS IBI-derived model, comparisons are
made to other CG water models in the literature, namely, the $k$-means based
potential of Hadley and McCabe\cite{hadley2010on} derived via the single state
(SS) IBI procedure (here referred to as the SS IBI potential) and the MARTINI
potential.\cite{marrink2007martini}
These models were chosen because they are short-ranged, non-polarizable, and 4:1
models.
For reference, these potentials are plotted in
Figure~\ref{fig:compare-potentials}.
Note that the MS IBI and SS IBI potentials are numerical (as they were derived
via IBI), while the MARTINI potential is represented by a 12-6 Lennard-Jones
potential with a well depth of 1.195~kcal/mol located at a separation of
5.276~\AA{}. 
Note that all of the potentials considered in this paper provide a close
estimate of the density of water at 1~atm and 305~K, as reported in
Table~\ref{tab:density}.
\begin{table}[ht]
  \ra{1.2}
  \setlength{\tabcolsep}{12pt}
  \centering
  \caption{Density of CG water at 305~K, 1~atm with various models.}
  \begin{tabular}{@{} l l @{}}
    \toprule
    Model & $\rho$, g/mL \\ \midrule
    TIP3P & 1.037 $\pm$ 0.004\\
    MS IBI & 1.027 $\pm$ 0.006\\
    SS IBI & 1.083 $\pm$ 0.008\\
    MARTINI & 1.015 $\pm$ 0.003 \\
    Chiu & 0.991 $\pm$ 0.003\\
    \bottomrule
  \end{tabular}
  \label{tab:density}
\end{table}

\begin{figure}[h!]
  \centering
  \includegraphics[width=3in]{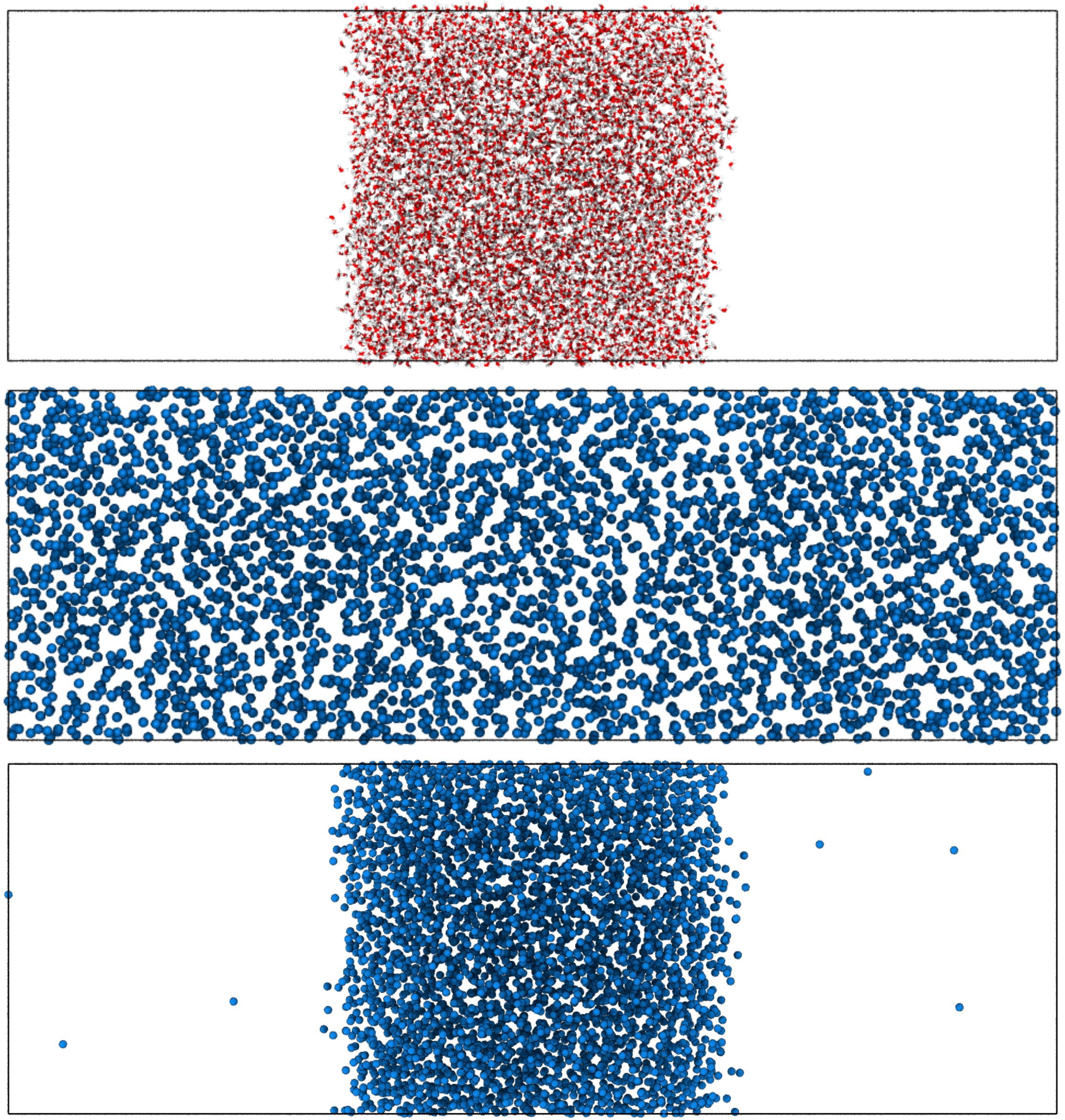}
  \caption{Simulation snapshots of droplets using the various models discussed.
  Top: all-atom TIP3P; middle: SS IBI; bottom: MS IBI. Atomistic and MS IBI
models agree, producing a system with a stable interface, whereas SS IBI does
not form a stable interface.}
  \label{fig:droplets}
\end{figure}

First considering the SS IBI potential, it can be seen that the well depth is
approximately 0.5~kcal/mol weaker than the MS IBI potential and shifted to
larger separations.
While this has little impact on the density or the structural correlations of
NVT and NPT states (not shown), given the potential was optimized to match these
RDFs, simulations of droplets show that the interfacial properties are not
sufficiently captured.
Specifically, as shown in Figure~\ref{fig:droplets}, simulations of atomistic
TIP3P, SS IBI, and MS IBI water were performed with interfaces.
From these it can be clearly seen that the SS IBI potential model fills the box,
rather than maintaining an interface.
In contrast, the MS IBI model maintains a stable interface in agreement with the
atomistic model. 
Thus, while an exact match to the experimental surface tension is not found for
the MS IBI potential, it is still sufficiently strong enough to maintain a clear
interface, providing a significant improvement over the SS IBI potential.

\begin{figure}[h!]
  \centering
  \includegraphics[trim={2.4in 0.7in 2.1in 1.2in},clip,width=3.0in]{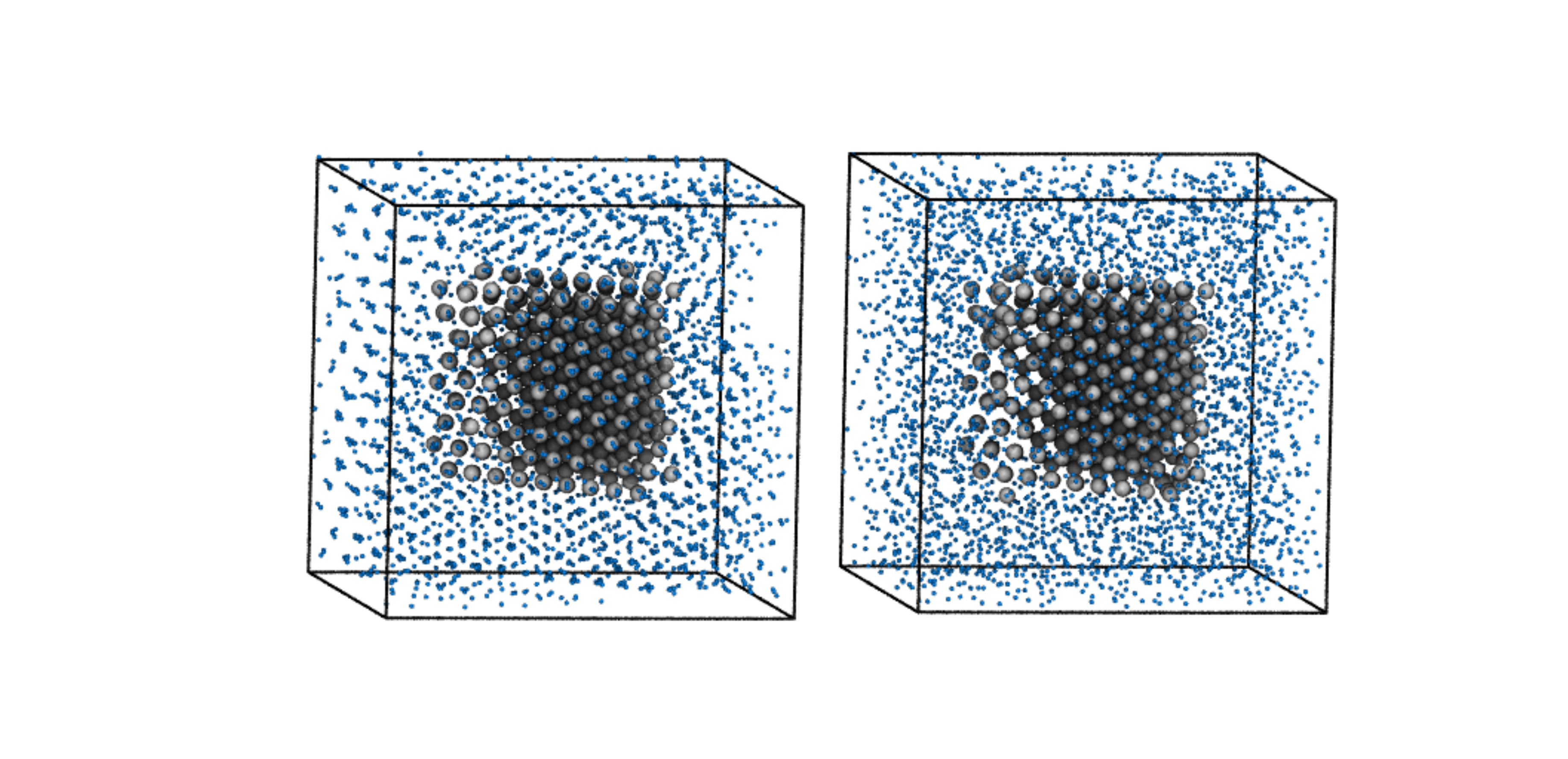}
  \caption{Configurations from simulations in the presence of a nucleation site
  with the MARTINI (left) and MS IBI (right) models. CG water beads colored
silver were kept fixed during the simulations, but were treated as the same type
as blue particles (i.e., the color is different to show the nucleation site).}
  \label{fig:crystal}
\end{figure}

\begin{figure}[h!]
  \centering
  \includegraphics{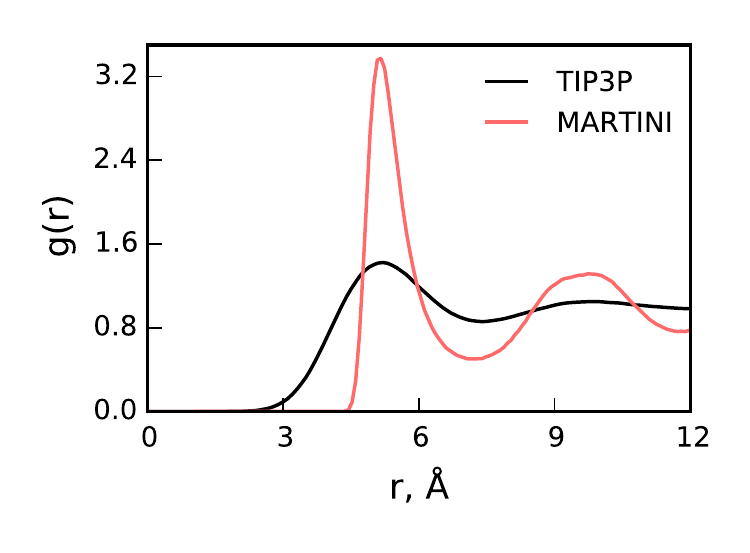}
  \caption{RDFs of the MARTINI model and the atomistic TIP3P model mapped to the
  CG level for the bulk NVT state.}
  \label{fig:martini-rdfs}
\end{figure}

It is also important that the potential is not so strong that the system can
solidify at physiological conditions.
For example, the MARTINI water model is known to spontaneously crystallize at
physiologically relevant temperatures.\cite{marrink2007martini}
This phenomenon is enhanced by the presence of interfaces (e.g., a lipid bilayer
surface), and requires the addition of unphysical ``antifreeze'' particles to
avoid crystallization.
While we note that modifications to the MARTINI water model exist (e.g., adding
charge polarization),\cite{yesylevskyy2010polarizable,zavadlav2015adaptive} only
the original MARTINI model was tested, since it more closely resembles the model
derived via MS IBI (i.e., both represent 4 water molecules as a single,
spherically symmetric interaction site).
To test the crystallization tendency, a nucleation site is generated with the
following protocol.
A crystalline state is generated by running a simulation with the MS IBI
potential in the NVT ensemble.
During this simulation, the temperature is decreased from 305~K to 1~K over
10~ns. 
A subsequent CG simulation is run at 1000~K, where the middle-most 1/8th of the
beads are kept fixed, resulting in a configuration that contains a crystal seed
surrounded by a fluid of CG water beads.
The beads in the crystal seed are kept fixed in the nucleation site simulations,
with interactions identical to the fluid interactions. 
While neither models shows a tendency to freeze at 305~K in the absence of a
nucleation site over 100 ns of simulation, the MARTINI model rapidly
crystallizes in the presence of a nucleation site, while the MS IBI potentials
remains fluid (\ref{fig:crystal}). 
Note, for a direct comparison with the MS IBI model derived here, antifreeze
particles were not used with the MARTINI model.
To ensure that the MS IBI system is not an amorphous solid structure, the ratio
of the diffusion coefficients with and without a nucleation site were calculated
for each model.
As shown in Table~\ref{tab:diffusion}, the diffusion coefficient of the MS IBI
potential model remains relatively unchanged when a nucleation site is added,
whereas a significant drop is seen for the MARTINI model, resulting from
crystallization.
Additionally, Figure~\ref{fig:martini-rdfs} plots the RDF of the MARTINI model for
the bulk NVT state as compared to the 4:1 mapped target data.
Clearly, the MARTINI potential does not accurately capture the structural
correlations of bulk water, further demonstrating the significant improvement of
the MS IBI model in reproducing key properties of water. 

\begin{table}[ht]
  \ra{1.2}
  \setlength{\tabcolsep}{12pt}
  \centering
  \caption{Ratio of diffusion coefficients from simulations with ($D_{nuc}$)
  and without ($D_{bulk}$) a nucleation site with different potentials. Diffusion
  coefficient $D$ calculated from the slope of a linear fit to the long-time
  MSD, using $MSD=6Dt$.} 
    
  \begin{tabular}{@{} l l @{}}
    \toprule
    Model & $D_{nuc}/D_{bulk}$ \\ \midrule
    MS IBI & 0.88 \\
    MARTINI & 0.02 \\
    \bottomrule
  \end{tabular}
  \label{tab:diffusion}
\end{table}

\section{Conclusions \label{sec:conclusion}}
In this work, the MS IBI method was used to derive the interactions for a 4:1
mapped CG water model. 
An improvement over previous models is made by simultaneously matching the fluid
structure to target data from bulk and interfacial states.
It was shown that a model that reproduces the structure and density of water
does not necessarily reproduce the interfacial properties and that the addition
of a droplet target state constrains the potential to also capture the
interfacial properties.
The resulting potential is able to accurately predict the density of water at
305~K at 1~atm, interfacial properties, and structural correlations.
Additionally, the model shows no tendency to spontaneously crystallize at
physiological conditions.
This is important, since inaccuracies in a water model can propagate as more
potentials are derived against it when simulating mixed systems.

This work highlights a key advantage of deriving potentials via the MS IBI
approach.
For simulations that cover multiple states, it is important to have a forcefield
that is accurate across the states of interest.
MS IBI allows this to be achieved by including target data from states that
represent structures present in the states of interest. 
This is realized here by including a multi-ensemble state to accurately model
the pressure-density relationship, and a droplet state to capture the
interfacial properties.
Another case where this would be beneficial is studying systems over multiple
phases, e.g., phase transitions in liquid crystals.
While clever approaches are taken to capture behavior across multiple
states,\cite{mukherjee2012derivation} a more systematic approach would be
useful.
Based on the results presented here, we foresee this method being useful for
deriving CG potentials for a wide range of applications.

\bibliography{main}
\end{document}